# Nanoflare Heating: Observations and Theory[1]

James A. Klimchuk
NASA-GSFC

Understanding how the solar corona is heated to multi-million degree temperatures, three orders of magnitude hotter than the underlying solar surface, remains one of the fundamental problems in space science. Excellent progress has been made in recent years, due in no small part to the outstanding observations from Hinode, but many important questions are still unanswered. The two long-standing categories of heating mechanisms---reconnection of stressed magnetic fields and dissipation of MHD waves---are both still under consideration. There is little doubt that both types of heating occur, and the real issue is their relative importance, which could vary from place to place on the Sun.

It is important to understand that reconnection heating and wave heating are both highly time-dependent (Klimchuk 2006; van Ballegooijen et al. 2011; Section 6.1). The timescale for energy release on a given magnetic field line is likely to be much less than a plasma cooling time, so we can consider the heating to be impulsive. The pertinent question is the frequency with which heating events repeat. If they repeat with a short delay, then the plasma is reheated before it experiences substantial cooling. This is call high-frequency heating, and will produce plasma conditions similar to steady heating if the frequency is sufficiently high. If the delay between successive events is long, the plasma cools fully before being reheated and the heating is considered to be low frequency.

Impulsive heating events are often called nanoflares. The meaning of the term is not always clear, however. Parker (1988) originally coined the name to describe a burst of magnetic reconnection in tangled magnetic fields. Low repetition frequency was assumed. Subsequently, many studies considered the hydrodynamic consequences of impulsive heating without specifying its cause, and it became convenient to adopt a generic term for any impulsive energy release on a small cross-field spatial scale, without regard to physical mechanism and without regard to frequency. Nanoflare started to be used in this way. That is the definition adopted here.

Another term with various meanings is "coronal loop." It sometimes refers to an observationally distinct feature in an image, assumed to coincide with a closed magnetic flux tube. A common misconception is that loops are much brighter than the background emission. In fact, they typically represent a small enhancement over the background of order 10% (Del Zanna & Mason 2003; Viall & Klimchuk 2011). Images often give a false impression because the color table assigns black to the minimum intensity, not to zero. Loops are useful to study because they can often be isolated from the background using a subtraction technique. It must be remembered, however, that they represent a small fraction of the coronal plasma and are, by definition, atypical. The diffuse component of the corona is in many ways more important and deserves greater attention than it has received.

The second definition of loop is more theoretical: a curved magnetic flux tube rooted in the photosphere at both ends, with approximately uniform plasma over a cross section. By this definition, the entire magnetically-close corona is filled with loops. I will use the term "strand" to

---

refer to the theoretical structure, and "loop" to refer to the observational feature. Loops are believed to be comprised of many thinner, unresolved strands.

This short chapter on observations and theory of nanoflares is nothing like an exhaustive review. Theoretical discussion is restricted to how the plasma evolves in response to a nanoflare. There is no attempt to discuss the theory of heating mechanisms. Citations are representative only and reflect a personal bias. Further information and additional references can be found in Klimchuk (2006, 2015).

1. Observational Discriminators

Some of the earliest evidence for low-frequency nanoflares came from the observation that warm (~1 MK) loops are over-dense compared to what is expected from steady heating; see the "coronal loops flowchart" (Klimchuk 2009) in the Proceedings of the Hinode 2 Workshop. More recently, coronal researchers have concentrated on four other observational discriminators of low-frequency and high-frequency heating: (1) intensity fluctuations, (2) time lags, (3) emission measure slope, and (4) very hot (>5 MK) plasma.

To understand these discriminators, it is helpful to review the characteristic response of a strand to impulsive heating. The panel at the top-left in Figure 1 shows the evolution of temperature (solid) and density (dashed) in a stand of $6 \times 10^9$ cm total length that is subjected to nanoflares of 100 s duration and 0.15 erg cm$^{-3}$ s$^{-1}$ amplitude (triangular heating profile). There is also a constant background heating of $10^{-5}$ erg cm$^{-3}$ s$^{-1}$. The nanoflares repeat every 3000 s, which is much longer than a cooling time, so this is in the low frequency regime. The simulation was performed with the EBTEL code (Klimchuk et al. 2008; Cargill et al. 2012), and only the last of several cycles is shown, when any influence of the initial conditions is gone. As is well understood, the plasma heats rapidly to high temperature due to the low density at the time of the nanoflare. The subsequent cooling is initially very rapid and dominated by thermal conduction. This then transitions into slower cooling that is dominated by radiation. Density rises during the conduction phase due to chromospheric "evaporation," and it falls during the radiation phase as plasma drains and "condenses" back onto the chromosphere. The peak in density, and therefore emission measure ($\propto n^2$), occurs well after the peak in temperature.

The panel on the top-right shows the same strand that is now heated by a quicker succession of weaker nanoflares. The amplitude is ten times smaller (0.015 erg cm$^{-3}$ s$^{-1}$) and the start-to-start delay is ten time shorter (300 s), so the time-averaged heating rate is the same. It corresponds to an energy flux through the footpoints of $7.5 \times 10^6$ erg cm$^{-2}$ s$^{-1}$, which is appropriate for active regions (Withbroe & Noyes 1977). Despite an equivalent time-averaged heating rate, the behavior is fundamentally different than the first case. Temperature and density now fluctuate about mean values of 3 MK and $3 \times 10^9$ cm$^{-3}$. Because the delay is much less than a cooling time, this is in the high-frequency regime.

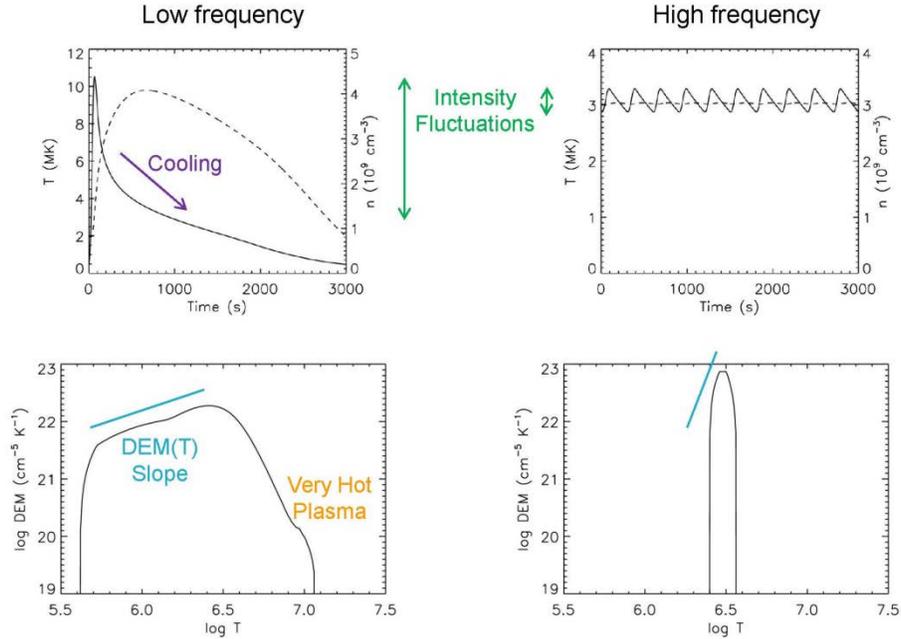

**Figure 1:** Top panels show the evolution of the strand-averaged coronal temperature (solid) and density (dashed) for low-frequency nanoflares (left) and high-frequency nanoflares (right). Bottom panels show the corresponding time-averaged differential emission measure distributions.

1.1 Intensity Fluctuations

Temporal variations in temperature and density produce temporal variations in emission, which can be used to detect nanoflares and measure their properties. The difficulty is that multiple events are observed together along the optically-thin line of sight. The composite light curve (intensity versus time) from many overlapping strands is nearly steady, even when the individual strands are highly variable. Sizable changes in intensity occur only when an unusually large event occurs or when there is a coherence in events of more typical size. For example, coronal loops are thought to be produced by "storms" of nanoflares, perhaps representing an avalanche process of some kind (e.g. Hood et al. 2016). Attempts have been made to count individual events, but these events are much larger than typical nanoflares, and estimation of their energy is fraught with uncertainty.

Although individual nanoflares are not generally detectable, their existence can be inferred from the composite emission from many unresolved events. Several approaches have been used. One indication of nanoflares is that the distributions of measured intensities are wider than expected from photon counting statistics if the plasma were slowly evolving (Katsukawa & Tsuneta 2001; Sakamoto et al. 2008). The distributions also have a skewed shape, as evidenced by small differences between the mean and median intensity (Terzo et al. 2011; Lopez Fuentes & Klimchuk 2016) and by the fact that the intensities are well represented by a log-normal distribution (Pauluhn & Solanki 2007; Bazarghan et al. 2008; see also Cadavid et al. 2016). Skewing of the distributions is expected from the exponential decrease in intensity as strands cool, which can also explain why Fourier power spectra are observed to have a power law form

(Cadavid et al. 2014; Ireland et al. 2015). Finally, the properties of observed light curves are consistent with impulsive heating (Tajfirouze et al. 2016).

1.2 Time Lags

If a cooling strand is observed with an instrument that can discriminate temperature (narrow band imager or spectrometer), the emission will peak first in the hottest channel and at progressively later times in cooler channels. Light curves with a clear hot-to-cool progression are typical of many coronal loops. What might seem surprising is that an unmistakable signature of cooling is present even in the nearly steady light curves characteristic of the diffuse corona. Viall & Klimchuk (2012) developed an automated procedure that measures the time lags between observing channels by cross correlating the light curves with varying temporal offset to see which offset maximizes the correlation. Using a combination of SDO/AIA observations and numerical simulations, they concluded that unresolved nanoflares of low to medium frequency are ubiquitous in the corona (Viall & Klimchuk, 2012, 2013, 2015, 2016, 2017; Bradshaw & Viall 2016). Note, however, that their results do not preclude the co-existence of high-frequency nanoflares along the same lines of sight.

The longest time delays found by Viall & Klimchuk exceed the predicted cooling times (Lionello et al. 2016; Section 7.4), though these delays tend to occur in the periphery of active regions, and longer strands are expected to cool more slowly. Also, uncertainties in the optically-thin radiative loss function must be taken into account. The measured delays could indicate a slow change in the envelope of nanoflare energies rather than the cooling if individual strands. Alternatively, they could be due to thermal nonequilibrium (Winebarger et al. 2016). This fascinating phenomenon occurs when steady (or high enough frequency) heating is strongly concentrated in the low corona. No equilibrium exists, and the stand experiences cycles of rising and falling temperature with periods of several hours (Antiochos & Klimchuk 1991). This is usually accompanied by the formation of a cold ($\sim 10^4$ K) condensation, which falls down one of the strand legs. While this is a likely explanation of coronal rain (Muller et al. 2004; Antolin et al. 2010) and of prominences (Antiochos et al. 1999; Karpen et al. 2003), Klimchuk et al. (2010) have argued that it is inconsistent with observations of coronal loops. It has recently been shown, however, that the condensation process can be aborted at modest ($\sim 1$ MK) temperatures (Mikic et al. 2013). Such behavior can explain long-period loop pulsations (Froment et al. 2015), which occur in isolated places in some active regions, but might also have more general applicability. Further study is needed.

1.3 Emission Measure Slope

A strand heated by low-frequency nanoflares experiences a wide range of temperatures during its evolution. The emission measure (EM) distribution is therefore very broad. In stark contrast, the EM distribution of a strand heated by high-frequency nanoflares is narrow. The lower panels in Figure 1 show the time-averaged differential emission measure distributions of the two examples (corona only; no transition region). The differential and regular emission measures are related according to EM($T$) = $T$*DEM($T$). The slope of the distribution coolward of the peak can be approximated by a power law and is a good indicator of nanoflare frequency. Low-frequency nanoflares produce smaller slopes than high-frequency nanoflares (Section 7.4; Bradshaw et al. 2012; Mulu-Moore et al. 2011; Warren et al. 2011). A wide range of slopes have been observed in active regions, indicating both low and high-frequency heating (Winebarger et al. 2011; Tripathi et al. 2011; Warren et al. 2012; Schmelz & Pathak 2012). The uncertainties are substantial, however (Guennou et al. 2013).

It has recently been shown that the range of slopes can be explained if nanoflares occur with a variety of energies and frequencies along the line of sight (Cargill 2014; Cargill et al. 2015; Lopez Fuentes & Klimchuk 2016). The distribution of frequencies must be broad and centered on an intermediate frequency in which the nanoflare delay is comparable to a cooling time (~1000 s). The EM slope will vary depending on the shift of the distribution toward higher or lower frequencies and possibly also on statistical fluctuations. It is important to note that these same distributions also reproduce the observed range of time lags (Bradshaw & Viall 2016). Figure 2, from a cellular automaton model of Lopez Fuentes & Klimchuk (2016), shows an example of a strand that is heated with a distribution of nanoflare frequencies and energies of the type advocated here.

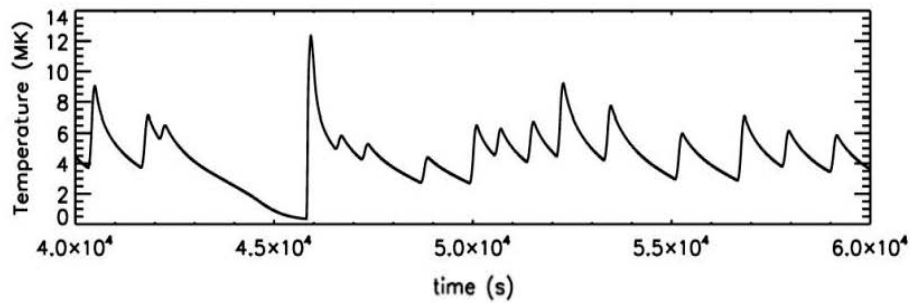

Figure 2: Temperature evolution of a strand heating by nanoflares from a cellular automaton model (after Lopez Fuentes & Klimchuk, 2016).

1.4 Very Hot (>5 MK) Plasma

The presence of very hot plasma in the corona is a strong indication of low-frequency nanoflares, since the heating rate needed to maintain steady plasma at such temperatures is extreme. For example, a strand of total length $L = 10^{10}$ cm requires an energy flux through the footpoints of $4 \times 10^8$ erg cm$^{-2}$ s$^{-1}$ to produce a steady apex temperature of 10 MK. If this energy were supplied by stressing of the field by footpoint motions, as envisioned by Parker, it would require continuous horizontal velocities in the photosphere of more than $10^6$ cm s$^{-1}$ (assuming an active region coronal field strength $B = 100$ G). This is more than an order of magnitude faster than observed. Higher temperatures and weaker fields would require even faster flows, since $v \propto T^{7/2}/(B^2L)$.

We refer to very hot plasma as the "smoking gun" of low-frequency nanoflares. Such plasma is difficult to observe, however, because it is expected to be very faint. As Figure 1 shows, the plasma cools rapidly and persists for only a short time. Its density is low because evaporation has not had time to fill the strand. Both factors contribute to a time-averaged emission measure that is very small. There have been multiple investigations to detect very hot plasma, most of them successful (Reale et al. 2009a,b; Ko et al. 2009; Patsourakos & Klimchuk 2009; Schmelz et al. 2009a,b; McTiernan 2009; Sylwester et al. 2010; O'Dwyer et al. 2011; Warren et al. 2011; Warren et al. 2012; Testa et al. 2011; Testa & Reale 2012; Teriaca et al.

2012; Ugarte Urra & Warren 2014; Del Zanna & Mason 2014; Caspi et al. 2015; Viall & Klimchuk 2017). Of particular note are the results from the EUNIS rocket spectrometer, which observed pervasive Fe XIX emission (~9 MK) in an active region (Brosius et al. 2014). Non-equilibrium ionization can further diminish the intensity of very hot spectral lines (Golub et al. 1989; Reale & Orlando 2008; Bradshaw & Klimchuk 2011), but such effects do not impact thermal bremsstrahlung emission observed in hard X-rays (Ishikawa et al. 2014; Hannah et al. 2016). Marsh et al. (2016) find that hard X-ray continuum spectra from the FOXSI sounding rocket and NuSTAR mission are consistent with low-frequency nanoflares.

In summary, the variety of different techniques for diagnosing coronal heating support the view that nanoflares occur with a wide range of energies and frequencies. Such a picture can reconcile observations that would otherwise seem to be contradictory. For example, Warren et al. (2012) measured the EM slopes in small subfields within 15 active regions and found a range of values indicating high-frequency heating in some cases and low-frequency heating in others. Viall & Klimchuk (2017) studied these same subfields using their time lag technique and found clear evidence of low and intermediate-frequency heating in every case, including those with steep slopes. All of the subfields also show evidence of very hot plasma, which can only come from low-frequency nanoflares. It seems that nanoflares of all frequencies are present, and that different techniques are sensitive to different parts of the frequency distribution.

Much more work needs to be done to determine how nanoflares are distributed in frequency and energy, and how these distributions vary in space and evolve with time. Among the important questions are the following? What causes the collective behavior responsible for loops? When does high-frequency heating persist for long enough to produce thermal nonequilibrium, with full or aborted condensations? What is the physical mechanism responsible for the heating?

I close by stressing the importance of studying very high temperature (>5 MK) emission. Such emission gives direct information on the energy-release process during low-frequency heating, when there is the least observational ambiguity. Much of the plasma at traditional coronal temperatures (~2 MK) has either cooled dramatically, in which case valuable information about the heating mechanism has been lost, or else has been evaporated from the chromosphere and is only an indirect bi-product of the heating. Emission line spectroscopy of very high temperature plasma is especially desirable. As already stressed, nanoflares are observed in aggregate due to line-of-sight overlap and finite spatial resolution. Only spectroscopy can sort out the properties (temperature, velocity, etc.) of non-uniform plasmas.